\documentclass[12pt]{article}

 \newenvironment{proof}
       {\medskip\noindent{\bf Proof.}}
       {\hfill$\Box$\medskip}

\newtheorem{defined}{Definition}

\newtheorem{exa}{Example}

\usepackage{graphicx}

\usepackage{thanks}
\usepackage{amsmath}

\usepackage{psfig}

\usepackage{url}
\usepackage{amssymb}

\usepackage{algorithm2e}

\newcommand{\myra}{\mbox{$\:\rightarrow\:$}}

\newcommand{\fa}{\mbox{$\forall$}}
\newcommand{\te}{\mbox{$\exists$}}

\newcommand{\NI}{\noindent}

\newcommand{\III}{\vspace{3 mm}}
\newcommand{\II}{\vspace{2 mm}}

\newcommand{\szkew}[1]{\relax \setbox0=\hbox{\kern -24pt $\displaystyle#1$\kern 0pt }%
\box0}
{\catcode`\@=11 \global\let\ifjusthvtest@=\iffalse}
\newcounter{oldmycaption}

\input{epsf}
\normalbaselines

\def\smallromani{\renewcommand{\theenumi}{\roman{enumi}}
\renewcommand{\labelenumi}{(\theenumi)}}

\let\oldmarginpar\marginpar
\renewcommand\marginpar[1]{\-\oldmarginpar[\raggedleft\footnotesize #1]%
{\raggedright\footnotesize #1}}

\title{A System for Distributed Mechanisms: \\
Design, Implementation and Applications
}
\author{Krzysztof R. Apt \thanksref{t1}\thanksref{t2} , Farhad Arbab\thanksref{t1}\thanksref{t3} and Huiye Ma\thanksref{t4}}
\thankstext{t1}{CWI, Science Park 123, 1098 XG Amsterdam, The Netherlands}
\thankstext{t2}{University of Amsterdam}
\thankstext{t3}{University of Leiden}
\thankstext{t4}{Technical University of Eindhoven}

\begin{document}
\maketitle

\begin{abstract}
  We describe here a structured system for distributed mechanism
  design appropriate for both Intranet and Internet applications. In
  our approach the players dynamically form a network in which they
  know neither their neighbours nor the size of the network and
  interact to jointly take decisions.  The only assumption concerning
  the underlying communication layer is that for each pair of
  processes there is a path of neighbours connecting them.  This
  allows us to deal with arbitrary network topologies.

  We also discuss the implementation of this system which consists of
  a sequence of layers.  The lower layers deal with the operations
  that implement the basic primitives of distributed computing, namely
  low level communication and distributed termination, while the upper
  layers use these primitives to implement high level communication
  among players, including broadcasting and multicasting, and
  distributed decision making.
% The lower layers deal with the operations
%   relevant for distributed computing only, while the upper layers deal
%   with the relevant aspects of the mechanism design, such as
%   computation of the desired decision and the taxes.  
This yields a highly flexible distributed system whose specific
applications are realized as instances of its top layer. This design
is implemented in Java.
  
  The system supports at various levels fault-tolerance and includes a
  provision for distributed policing the purpose of which is to
  exclude `dishonest' players.  Also, it can be used for repeated
  creation of dynamically formed networks of players interested in a
  joint decision making implemented by means of a tax-based mechanism.
  We illustrate its flexibility by discussing a number of implemented
  examples.

%\footnote{Initial, short version of this paper appeared as \cite{AAM08}.}
\end{abstract}
%  \category{C.2.4}{Distributed Systems}{Distributed applications}

%  \category{J.4}{Social and Behavioral Sciences}{Economics}

%  \terms{Systems, Economic Applications}

%  \keywords{mechanism design, distributed system, fault-tolerance, distributed policing}

% \begin{bottomstuff}
% Authors' addresses: 
% Krzysztof R. Apt, CWI, Science Park 123, 1098 XG Amsterdam,
% The Netherlands and
% ILLC, University of Amsterdam.
% Farhad Arbab, CWI and University of Leiden, The Netherlands.
% Huiye Ma, Technical University of Eindhoven, The Netherlands.

% The work of Huiye Ma was funded by the NWO project
% DIACoDeM, No 642.066.604.
% \end{bottomstuff}

%-------------------------------------------------------------------------
\section{Introduction}
\label{sec:intro}

\subsection{Background and motivation}

Mechanism design is one of the important areas of economics.  To quote
from \cite{Eco07}, it deals with the problem of `how to arrange our
economic interactions so that, when everyone behaves in a
self-interested manner, the result is something we all like.'  So
these interactions are supposed to yield desired social decisions when
each agent is interested in maximizing only his own utility.

The traditional approaches rely on the existence of a
central authority who collects the information from the players,
computes the decision and informs the players about the outcome and
their taxes. The increasing reliance on decision making carried out through
Internet leads to a natural need for distributed solutions that do not
rely on any central authority. But how to translate such real-world
considerations into design and implementation that can be used in
practice?

This question was recently addressed in a series of papers on
distributed mechanism design.  In this setting no central authority
exists and the decisions are taken by the players themselves, so they
need to communicate with each other to take jointly the desired
decision and to compute the taxes.  The challenge is to appropriately
combine the techniques of distributed computing with those that deal
with the matters specific to mechanism design, notably rationality
(i.e., appropriately defined self-interest) and truth-telling (i.e.,
incentive compatibility).

However, to properly implement decision making in the context of the
Internet one needs to address other issues, as well.  First of all,
one should provide an \emph{open system} that can deal with the
initially unknown number of interested users.  Second, one should
support \emph{connectivity} between the users who can dynamically join
the system.  Further, one should be able to cope with unreliable
(hacked or faulty software or hardware in) user devices that can lead
to system failures. Also, it is desirable to provide ways of dealing
with the dishonest users, such as their identification and possible
exclusion.  These additional issues have been hardly considered in the
papers of distributed mechanism design.  In this paper we discuss a
system for distributed mechanism design that addresses all these
issues and that can be readily used both in the Intranet and the
Internet setting.

To our knowledge, no working implementation of a distributed mechanism
design has been described in the literature.  The gap between
published distributed mechanism designs and a working distributed
implementation of such systems cannot be filled simply by applying
individual techniques, such as cryptography, digital signatures,
fault-tolerance methods, etc. To support a large class of distributed
decision making mechanisms over the Internet and Intranets, a coherent
distributed platform is needed wherein state-of-the-art cryptography,
fault-tolerance, etc., tools can be employed, as necessary. One of our
contributions is to bring known useful models and techniques from
distributed computing and software engineering to bear on the
distributed implementation of distributed mechanisms.

\subsection{Related work}

A number of recent papers deal with different aspects of distributed
computing in connection with game theory and mechanism design.

Among them, some focus on complexity such as communication complexity.
Some target computation/communication/incentive compatibility and
eventually faithful implementation. Others try to build a secure
computation in a distributed system. More recently, there has been a
series of work on distributed constraint optimization and partial
centralized techniques.

The authors of \cite{Monderer1999D} focused on message communication
by players in a distributed game. However, they assume that there is a
center to which every player is directly connected. An influential
paper \cite{Feigenbaum2002D} introduced the notion of distributed
algorithmic mechanism design emphasizing the issues of computational
complexity and incentive compatibility in distributed computing.
Next, \cite{PS04} studied the distributed implementations of the 
Vickrey-Clarke-Groves (in short VCG) mechanism.
However, in their approach there is still a center that is
ultimately responsible for selecting and enforcing the outcome.

The authors of \cite{Shneidman2004S} considered the problem of
creating distributed system specifications that will be faithfully
implemented in networks with rational (i.e., self-interested) nodes so
that no node will choose to deviate from the specification.  They used
interdomain routing as an example and suggested ways to detect when
nodes deviate from their specified communication.  In turn,
\cite{Izmalkov2005R} proposed in the context of secure computation a
stronger form of computation in that it solely depends on players'
rationality instead of their honesty.

% In \cite{Petcu2005D} the authors present a new, complete method for
% distributed constraint optimization, based on dynamic programming. It
% is based on a so-called utility propagation method, inspired by the
% sum-product algorithm which is correct only for tree-shaped constraint
% networks. They show how to extend this algorithm to arbitrary
% topologies.
Researchers of \cite{Petcu2006M} introduced the first distributed
implementation of the VCG mechanism.  The only central authority
required was a bank process that collects the payments and is in
charge of the computation of taxes.  The authors also discussed a
method to redistribute some of the VCG payments back to players.  This
paper is most closely related to our work and we discuss it more fully
in the next subsection.  Finally, \cite{Petcu2007P} proposed a new
partial centralization technique, PC-DPOP, based on the DPOP algorithm
of \cite{Petcu2005D}.  PC-DPOP provides a better control over what
parts of the problem are centralized and allows this centralization to
be optimal with respect to the chosen communication structure.

\subsection{Contributions}

In this paper we propose a platform for distributed mechanism design
that can be readily used both in the context of the Intranet and
Internet and customized to specific purposes. Also it can be used for
a repeated distributed decision making process, each round involving a
different group of interested players.

Our platform supports the distributed implementation of the large
class of tax-based mechanisms that implement the decisions either in
dominant strategies or in an ex post Nash equilibrium (see, e.g.,
\cite{MWG95}).  This aspect of our work is closest to
\cite{Petcu2006M} whose approach is based on distributed constraint
programming.  In contrast, our approach builds upon a very general
view of distributed programming, an area that developed a variety of
techniques appropriate for the problem at hand and that is more
appropriate for Internet applications.  

The main conceptual difference between our work and \cite{Petcu2006M}
is that we propose a more generic design that consists of a sequence
of layers by means of which secure communication in arbitrary network
topologies including ring, tree, forest, and graph can be supported
and on top of which a larger range of mechanisms can be run. In
contrast, in \cite{Petcu2006M}, a player is randomly chosen and the
players are organized in a tree with the chosen player at its root.
This allows one to minimize the number of messages needed to compute
the decision and the taxes but makes the chosen player a vulnerable
part of the system.

To support the Internet-based applications we ensure connectivity
between the players by means of a \textbf{backbone} of interconnected
gateways and \textbf{local registries}.  The dynamic network
creation is realized by means of players' \textbf{registration} in
their local registries.  This allows us to support the open system
aspect of the platform by allowing the presence of initially unknown
number of interested players.

To realize concrete applications we only need to provide a backbone of
local registries and select specific registration schemes for
participating in the mechanism.  The former can be taken care of by
stipulating that each geographic or logical region, such as a country,
city, or Internet domain has its own local registry.  Interested
players can find the addresses of their respective local registries in
public fora, e.g., local government web sites.  To take care of the
latter we can for example stipulate that the registration is
successful only if it took place before a certain deadline that refers
to a clock in player's region, or if some quorum (minimum number) of
registered players is reached at each local registry, and/or if a
global quorum of registered players is reached.

Our platform is built out of a number of layers.  This leads to a
flexible, hierarchical design in which the lower layers are concerned
only with the matters relevant for distributed computing, such as
communication and synchronization issues, and are clearly separated
from the upper layers that deal with the relevant aspects of the
mechanism design, such as computation of the desired decision.

More specifically, the lowest communication layer allows us to detect
process failure and provides an asynchronous, non-order-preserving
\texttt{send} operation.  The next layer provides a message efficient,
fault-tolerant distributed termination detection (see, e.g.,
\cite{MC98}) algorithm. In turn, the high-level communication layer
supports a generic \textbf{broadcast command} that supports
communication among players and ensures that each broadcast message is
eventually delivered to each registered player. Its implementation
relies only on the assumption that for each pair of registered players
there is a path of neighbouring processes connecting them.  Any
specific application, such as an appropriate instance of the Groves
mechanism (see, e.g., \cite{MWG95}), is realized simply as an
\emph{instantiation of a top layer}.  Finally, the deliberately
limited GUI prevents players from tampering with the system. In fact,
the players can use the GUI only to register and to submit their types.

This layered architecture, in conjunction with the use of local
registries and registration requirement, offers a number of novel
features and improvements to the approach of \cite{Petcu2006M}, to wit

\begin{itemize}

\item we deal with a larger class of mechanisms, notably Groves
  mechanisms. They include the VCG mechanism, various forms of
  redistributions of VCG payments recently studied in the literature
  (and considered in \cite{Petcu2006M}), and the Groves mechanism
  concerned with the problem of buying a path in a network, introduced
  in \cite{NR01}.  Additionally, we can easily tailor our platform to
  other tax-based mechanisms, such as Walker mechanism (see
  \cite{Wal81}),

\item we support open systems in which the number of players can be unknown,
  
\item the above-mentioned bank process of \cite{Petcu2006M}  is replaced by a weaker
  \textbf{tax collector} process. It is needed only for the mechanisms
  that are not budget balanced, wherein it is a passive process used
  only to receive messages in which some players either report taxes
  they need to pay (when the mechanism is feasible) or submit
  financial claims (when the mechanism is not feasible),
  
\item \textbf{fault-tolerance} is supported at various levels,
  including the message transmission level and the player processes
  level (with an option for a \emph{restart} in the case of the
  detection of a failed player process),

\item a multi-level protection against \textbf{manipulations} is provided,

\item our platform makes it possible to implement \textbf{distributed policing}
  that provides an alternative to a `central enforcer' whose
  responsibility is to implement the outcome decided by the agents and
  collect the taxes (see, e.g., \cite[ page 366]{FSS07}).

\end{itemize}

Fault-tolerance at the mechanism design level means that the final
decision and taxes can be computed even after some of the processes
that broadcast the players' types crash: the other processes then
still can proceed. Observe that detecting a non-responsive or
unreachable process as a fault and recovering the system to allow the
rest of the processes to proceed is the best that can be expected from
the platform layers.  What the top layer processes do in reaction to a
detected fault, whether they can reuse any of the partial results
already computed in the running mechanism, or have to restart, etc.,
depends on the type of the application they are engaged in and its
specific rules. Fault tolerance is achieved by a provision of failure
detection (in the BTTF layer) and by the duplication of the
computation by all players.

Such a redundancy is common in all approaches to fault-tolerance (and
also used to prevent manipulations, see \cite[ page 366]{FSS07}). In
\cite{Shneidman2003U} it is used to realize two natural requirements
for a distributed mechanism implementation: computation compatibility
and communication compatibility.  Redundancy was intentionally avoided
in \cite{Petcu2006M} which aimed at minimizing the overall
communication and computation costs. In our approach it allows the
fastest process to `dominate' the computation and move it forward more
quickly.

This design is implemented in Java and was tested on a number of
examples including Vickrey auction with redistribution, unit demand
and single-minded auctions, the problem of buying a path in a network,
and a sequential mechanism design, described briefly in the second
part of the paper and more fully in Appendix C.

\subsection{Paper organization}

In the next section we review the basic facts about the tax-based
mechanisms, notably the Groves family of mechanisms. Then in Section
\ref{sec:our} we discuss the issues that need to be taken care of when
moving from the centralized tax-based mechanisms to distributed ones
and what approach we took to tackle these issues. The details of our
design and implementation are provided in Section \ref{sec:details}.

Next, in Section \ref{sec:manip}, we discuss four important advantages
of our design: security, distributed policing and fault-tolerance and
the simple way of implementing repeated mechanisms.  The aim of
Section \ref{sec:examples} is to present a number of examples of
mechanisms that we implemented using our system. Then, in Section
\ref{sec:conc} we provide conclusions and discuss future work.
Finally, in Appendix A we discuss the details of our algorithm that
computes the tax scheme and in Appendices B and C we
present a sample interaction with our system and a more detailed
account of the implemented examples.

\section{Mechanism design: the classical view}
\label{sec:classical}

We recall here briefly tax-based mechanisms, notably the family of Groves mechanisms,
see, e.g., \cite[ Chapter 23]{MWG95}.
Assume a set of \textbf{decisions} $D$, a set $\{1, \ldots, n\}$ of players,
for each player a set of \textbf{types} $\Theta_i$ and a \textbf{utility function}
$
v_i  : D \times \Theta_i \rightarrow \mathbb{R}.
$
In this context a type is some private information known only to the player, for example
a vector of player's valuations of the items for sale in a multi-unit auction.

A \textbf{decision rule} is a function $f: \Theta \rightarrow D$,
where $\Theta := \Theta_1 \times \cdots \times \Theta_n$.  A decision
rule $f$ is called \textbf{efficient} if for all $\theta \in \Theta$
and $d' \in D$
\[
\sum_{i = 1}^{n} v_i(f(\theta), \theta_i) \geq \sum_{i = 1}^{n} v_i(d', \theta_i),
\]
and \textbf{strategy-proof} 
if for all $\theta \in \Theta$, $i \in \{1,\ldots,n\}$ and
$\theta'_i \in \Theta_i$
\[
v_i(f(\theta_i, \theta_{-i}), \theta_i) \geq
v_i(f(\theta'_i, \theta_{-i}), \theta_i),
\]
where $\theta_{-i} := (\theta_1, \ldots, \theta_{i-1}, \theta_{i+1}, \ldots, \theta_n)$ and
$(\theta'_i, \theta_{-i}) := (\theta_1, \ldots, \theta_{i-1}, \theta'_i, \theta_{i+1},$ $\ldots, \theta_n)$.

In mechanism design one is interested in the ways of inducing the
players to announce their true types, i.e., in transforming the
decision rules to the ones that are strategy-proof. The resulting
mechanisms are called \textbf{incentive compatible}.  In
\textbf{tax-based} mechanisms (called \textbf{direct mechanisms}
in the economics literature) this is achieved by extending
the original decision rule by means of \textbf{taxes} that
are computed by a central authority from the vector of the received
types, using players' utility functions.

We call the tuple
\[
(D, \Theta_1, \ldots, \Theta_n, v_1, \ldots, v_n, f)
\]
a \textbf{decision problem}. Given a decision problem, in the classical setting,
one considers then the following sequence of events, where $f$ is a
given, publicly known, decision rule:
\begin{enumerate} \smallromani

\item each player $i$ receives a type $\theta_i$,

\item each player $i$ announces to \emph{the central authority} a type $\theta'_i$;
this yields a joint type $\theta' := (\theta'_1, \ldots, \theta'_n)$,

\item the central authority then makes the decision $d := f(\theta')$,
computes the sequence of taxes $t := g(\theta')$, where
$g : \Theta \rightarrow \mathbb{R}^n$ is
a given function, and communicates to each player $i$ the decision $d$ and the tax $|t_i|$ he
needs to pay to (if $t_i \leq 0$)
or to receive from (if $t_i > 0$) the central authority.

\item the resulting utility for player $i$ is then
$u_i(d,t) := v_i(d, \theta_i) + t_i$.
\end{enumerate}

Each \textbf{Groves mechanism} is obtained using $g(\theta') :=
(t_1(\theta'), \ldots, t_n(\theta')), $ where for all $i \in \{1,
\ldots, n\}$

\begin{itemize}
\item
$
h_i: \Theta_{-i} \rightarrow \mathbb{R}
$
is an arbitrary function,

\item $t_i : \Theta \rightarrow \mathbb{R}$ is defined by\footnote{$\sum_{j\not=i}$ is a
shorthand for the summation over all $j \in \{1,\ldots,n\}, \ j
\not=i$.}
\[
t_i(\theta') :=
  h_i(\theta'_{-i}) + \sum_{j \neq i} v_j(f(\theta'), \theta'_j).
\]
\end{itemize}
Intuitively, the sum $\sum_{j \neq i} v_j(f(\theta'), \theta'_j)$ represents the
society benefit from the decision $f(\theta')$, with player $i$ excluded.

The importance of the Groves mechanisms is revealed by the following crucial result,
in which we refer to the expanded decision rule $(f, g): \Theta \rightarrow D \times \mathbb{R}^n$.
\III

\NI
\textbf{Groves Theorem}
Suppose the decision rule $f$ is efficient. Then in each Groves mechanism
the decision rule $(f,g)$ is strategy-proof w.r.t.~the utility functions $u_1, \ldots, u_n$.
\III

The proof is remarkably straightforward so we reproduce it for the convenience of the reader.
%%\II

%%\NI
\begin{proof}
Since $f$ is efficient, for all $\theta \in \Theta$, $i \in \{1, \ldots, n\}$ and
$\theta'_i \in \Theta_i$ we have

\begin{align*}
u_i((f,g)(\theta_i, \theta_{-i}), \theta_i) &=
\sum_{j=1}^{n} v_j(f(\theta_i, \theta_{-i}), \theta_j) +  h_i(\theta_{-i})\\
&\geq \sum_{j=1}^{n} v_j(f(\theta'_i, \theta_{-i}), \theta_j) +  h_i(\theta_{-i}) \\
& = u_i((f,g)(\theta'_i, \theta_{-i}), \theta_i).
\end{align*}
\end{proof}
%%\HB
%%\VV

When for a given tax-based mechanism for all $\theta'$ we have $\sum_{i = 1}^{n} t_i(\theta') \leq 0$,
the mechanism is called \textbf{feasible} (which means that it can be realized without external
financing) and when for all $\theta'$ we have $\sum_{i = 1}^{n} t_i(\theta') = 0$,
the mechanism is called \textbf{budget balanced} (which means that it can be realized without a deficit).

Each Groves mechanism depends on the functions $h_{1}, \ldots$, $h_{n}$.
A special case, called \textbf{Clarke mechanism}, or \textbf{\textit{Vickrey-Clarke-Groves mechanism}}
(in short \textbf{VCG}) is obtained by
using
\[
h_i(\theta'_{-i}) := - \max_{d \in D} \sum_{j \neq i} v_j(d, \theta'_j).
\]

%\NI
So then

\[
t_i(\theta')  := \sum_{j \neq i} v_j(f(\theta'), \theta'_j) - \max_{d \in D} \sum_{j \neq i} v_j(d, \theta'_j).
\]

Hence for all $\theta'$ and $i \in \{1,\ldots,n\}$ we have
$t_i(\theta') \leq 0$, which means that the VCG mechanism is feasible
and that each player needs to make the payment $|t_i(\theta')|$ to the
central authority.  Other feasible Groves mechanisms exist in which
some players receive payments and others have to make payments, for
example the one proposed in ~\cite{Cav06}.  Yet other Groves
mechanisms are not feasible, for example the one concerned with buying
a path in a network, due to \cite{NR01}.  In that mechanism the
players with non-zero taxes need to receive the payments from 
central authority.  We discuss these examples in Subsection
\ref{subsec:Groves}.

\section{Our approach}
\label{sec:our}

In our approach we relax a number of the assumptions made when
introducing mechanism design.  More specifically we assume that

\begin{itemize}
\item there is no central authority,

\item players interested in participating in a specific mechanism register to join an open system wherein that
mechanism runs. A tax collector process is part of this system,

\item the players whose registration is accepted inform other registered players about
their types,

\item once a registered player learns that he has received the types from all registered players, he computes
the decision and the taxes, sends this information to other registered players, and possibly the tax collector process,
and terminates his computation.

\end{itemize}

We also assume that there is no collusion among the players.
This leads to an implementation of the mechanism design by means of
anonymous (i.e., name independent) distributed processes, in absence
of any central authority.  Because of the distributed nature of this
approach no global state, in particular no global clock, exists.  The
computation of the decision and of the taxes is carried out by the
players themselves.

% In this revised setting the following sequence of events takes place for each player $i$:

% \begin{enumerate} \smallromani
% \item he receives a type $\theta_i$,

% \item he decides whether to register,

% \item if he registers and is admitted, he announces to \emph{other players} a type $\theta'_i$;

% \item after he receives all types $\theta' := (\theta'_1, \ldots, \theta'_n)$ from players who
% registered he computes the decision $d := f(\theta')$ and the sequence of taxes $t := g(\theta')$
% and communicates to each player $j$ the decision $d$ and the tax scheme $tax(t_j)$.
% \end{enumerate}

\subsection{Player processes and local registries}

As it stands, this revised setting is not clear on a number of counts.
First, we need to clarify the registration process, in particular what
it implies and when it ends.  In our approach each player is
represented by a process, in short a \textbf{player process}.  A
player who wishes to join a specific mechanism (e.g., an auction) must
register with a \textbf{local registry}.
% Each geographic or logical
% region, such as a country, city, or Internet domain can have its own
% local registry.  Players can find the addresses of their respective
% local registries in public fora, e.g., local government web sites.
Local registries are linked together in a network that satisfies the
full reachability condition described in Subsection \ref{subsec:BTTF}
(and we assume one of them is designated as the initiator mentioned in
that subsection).  Receiving his registration request, a local
registry verifies the eligibility of a player (e.g., whether his IP
address puts him under the jurisdiction of this registry) and accepts
his request if the registration conditions for the specific mechanism
(e.g., a deadline) are met.

\subsection{Generic \textbf{broadcast} command}

Second, once the registration process ends, in the resulting network a
player process may not know the identities of other player processes,
so the announcement of one's type to all other players needs to be
explained. In our approach we assume that once a player process is
registered, it joins the network of (registry and player) processes
wherein a generic \textbf{broadcast} command is available. The
implementation of this command relies only on the assumption that for
each pair of players there is a path of neighbouring processes
connecting them.  This allows us to deal with arbitrary network
topologies in a simple way.

The topology of this network is irrelevant both from the point of view
of the individual processes, as well as the semantics of the broadcast
command.  The full reachability of the backbone network of local
registries is enough to ensure that as long as each player process
knows and is known by its local registry, full reachability also holds
for the whole network.  The broadcast command uses the connectivity of
this network to ensure that a copy of a broadcast message is
eventually delivered to every registered player in finite time.  These
messages are transmitted through paths managed in a lower layer which
the player processes \emph{cannot access}.

This automatically prevents manipulation by player processes of
messages originating from or destined for other players.  Such
manipulations are possible in other schemes, such as in \cite[ page
366]{FSS07}, where the player processes connected in a ring are
computing a Vickrey (second-price) auction of a single good. The
processes are expected to pass around a message containing the top two
bids for that good. This opens the possibility of cheating by a process
by simply manipulating the messages that it is expected to pass through.
Indeed, by putting a high
bid and by substantially lowering the second lowest bid a player
process can get the good more cheaply (at least when it is the last
player process to bid). In our set up a player process cannot
access the bids of other player processes before broadcasting its own
bid (unless one explicitly considers a sequential set up, see
Subsection \ref{subsec:nonGroves}).  Moreover, no messages destined for a
player process pass through another player process.

\subsection{Distributed termination detection}

Third, we need to clarify how each player process will know that he indeed
received the types announced by \emph{all other} registered players.  We solve
this problem by assuming that each player process
after broadcasting the player's type
participates in a \textbf{distributed termination detection
    algorithm} the aim of which is to learn whether all players have
indeed broadcast their types.  This algorithm is tailored to deal with
the communication by means of multicasting (which subsumes
broadcasting).

If this algorithm detects termination, the player process
knows that he indeed received all types, and in particular can determine at
this stage the number of players and their (alias) names.
From that moment on each player process uses the same naming scheme
when referring to other player processes.  The uniqueness is ensured by a local
scheme for generating globally unique player identifiers.
More generally, we use the distributed termination detection algorithm to delineate
the end of each \emph{phase} of the distributed computation: registration, type broadcast, etc.,
i.e., for \textbf{barrier synchronization} (see, e.g., \cite{And05}).

This allows us to deal with the 
\textbf{repeated mechanisms} that involve several rounds of
decision making by means of the same given mechanism,
each time involving a possibly different group of players.  To this
end we need to logically separate each round of the mechanism.  This
is handled, using our distributed termination detection
algorithm for barrier synchronization.

\subsection{Tax schemes}

Fourth, to ensure the correctness of the above approach, it is crucial
that each player process computes the same decision and the same
information concerning taxes.  The former is taken care of by the fact
that each player process uses the same, publicly known, decision rule
$f$ that each player learns, for example from a public bulletin board,
and that is used by the player process after its registration is accepted.

Further, each player process applies $f$ to the same input $\theta'$
and computes \emph{the same} \textbf{tax scheme} by which we mean a
specific vector of payments $tax(t_1), \ldots tax(t_n)$ computed from
the tax vector $(t_1, \ldots, t_n)$, where $tax(t_j)$ specifies the
amounts that player $j$ has to pay to other players and possibly the
tax collector from his tax $t_j$.  All tax schemes $tax(t_1), \ldots
tax(t_n)$ then determine `who pays how much to whom'.  In general most
taxes equal 0, so we optimize the computation by generating
\textbf{reduced tax schemes} in which only non-zero entries are listed
and by multicasting them instead of broadcasting.  

Note also that to compute the taxes each player process needs to know
the utility functions of other player processes. This simply means
that each player knows that all the players are participating in the
same mechanism (for instance unit demand auction).  The tax collector
process is only needed for the mechanisms that are not budget
balanced.

\section{Implementation}
\label{sec:details}

\begin{figure}[htbp]
%\centerline{\psfig{figure=newlayers2.eps,height=5cm}}
\centerline{\includegraphics[height=5cm]{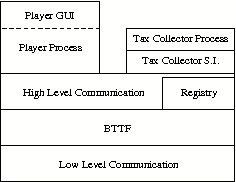}}
\caption{\label{F.Layers} Implementation architecture}
\end{figure}

Our distributed mechanism design system is implemented in Java using
threads and sockets. The implementation follows the guidelines
explained in the previous section.  Figure~\ref{F.Layers} shows the
overall architecture of our system and the different layers of
software used in its implementation. 

The first two layers provide support for asynchronous communication
and an appropriate fault-tolerant distributed termination
detection algorithm for arbitrary network topologies with the
asynchronous broadcasting as the communication primitive. Their
implementation is about 9K lines of Java code. The implementation of
the remainder of the system (developed by the third author) is about
4K lines of Java code.  Each entity in this architecture communicates,
either through function calls or method invocations, \emph{only} with
its adjacent entities.  Specific applications are realized by
instantiating the crucial player process layer.

\subsection{Low Level Communication}
\label{subsec:LLC}

The Low Level Communication (LLC) is a thin layer that supports (1)
locally generated, globally unique process identifiers, and (2)
reliable non-order-preserving, asynchronous, targeted communication,
exclusively through the exchange of passive messages between
processes.\footnote{This layer consists of just some 700 lines of code
  and any low-level message passing mechanism that provides
  asynchronous, non-order-preserving send and receive can be used
  instead of it.}  The only means of communication between processes
in LLC is through message passing, where no transfer of control takes
place when messages are exchanged.
%
% \emph{Targeted} means that the sender of a message explicitly specifies the
% recipient of the message. \emph{Asynchronous} means that the receiver of
% a message is not guaranteed to have received the message upon the
% completion of the send operation. \emph{Non-order-preserving} means that
% the temporal order of messages sent from the same sender to the same
% receiver is not necessarily preserved.  \emph{Reliable} means that every message
% sent by a sender will be received by its target receiver in finite but
% indeterminate time, without alteration and in its entirety (unless the
% receiver process fails or terminates).  
A message sent to a non-existent
(terminated or failed) process will be returned to its sender intact,
in finite but indeterminate time (a time-out).

The interface provided by the LLC layer contains the two operations {\tt
llsend(m, r)} and {\tt llreceive(m, t)}.  The {\tt llsend(m, r)} operation
sends the message {\tt m} to its target process {\tt r} and returns a
Boolean value that indicates the success or failure of the operation.
A send operation may fail, for instance, if the size of the message is
above the capacity threshold of the transport mechanism.
%, or due to other possible internal errors. 
Successful send simply means that
the message has been dispatched on its way to its specified target.

The {\tt llreceive(m, t)} operation blocks its calling process,
$p$, until either (a) a message sent to $p$ has arrived, or (b) the
specified time-out {\tt t} has expired.  In the first case, {\tt
llreceive()} returns {\tt true} and passes the received message
in {\tt m}.  In the second case, this function returns {\tt
false} to indicate that the time-out {\tt t} has expired.
% For
% convenience, we use the shorthand {\tt llreceive(m)} for the
% common situation where the time-out {\tt t} is infinity.

\subsection{BTTF}
\label{subsec:BTTF}

The Back To The Future (BTTF) layer implements a message efficient,
fault-tolerant distributed termination detection (DTD) algorithm, on
top of the LLC layer. The details of the BTTF DTD algorithm lie beyond
the scope of this paper and will be described elsewhere.

The DTD functionality provided by the BTTF layer can be used for
barrier synchronization as well as for termination detection.  
Once
%{\tt passiveReceive()} indicates 
termination has been detected, the
calling process knows that all processes in the system have reached
the same `termination barrier'.  This termination barrier is either
the actual termination of the processes, or the virtual termination of
only the current phase of the activity in the system.  In the first
case, the calling process must perform its local clean-up and
terminate.  In the second case, the process must start a new {\em
  phase} of its computation by calling the initialization function of
the BTTF layer once more.

The implementation of the BTTF layer requires only the {\tt llsend(m,
r)} and {\tt llreceive(m, t)} operations provided by the LLC layer.
It provides an interface that supports the following functions:

\begin{itemize}
\item
{\tt initializeBTTFWave($\ldots$)}
This function initializes the calling process, enabling it to
participate in the global computation.  The
details of the parameters of this function are beyond the scope
of this paper.

% \item
% {\tt insertBuddy(p)}
% This function call inserts the specified process {\tt p} in the
% buddies set of the calling process.

% \item
% {\tt removeBuddy(p)}
% This function call removes the specified process {\tt p} from the
% buddies set of the calling process.

\item
{\tt send(m, T)}
This function implements a delayed multicast operation.
It schedules a copy of the message {\tt m} to be sent
to every process in the target set of processes {\tt T}.
The actual dispatch of the messages to their specified targets will take
place upon a subsequent call to one of the functions
{\tt prioritySend()},
{\tt receive()}, or
{\tt passiveReceive()}.

\item
{\tt prioritySend(m, T)}
This function implements a multicast operation.
It first sends all messages scheduled by earlier calls to {\tt
send()}, if any, and then sends a copy of the message {\tt m} to
every process in the target set of processes {\tt T}.

\item
{\tt receive(m, t)}
The parameter {\tt t} is an integer value.  Negative {\tt t} values
indicate indefinite wait, and non-negative values specify a time-out value in
milliseconds.
A call to this function blocks until either the specified
time-out expires, or a message sent to the calling process
is available.
If the specified time-out expires, the return result of this
function is {\tt false} and the value of {\tt m} is undefined.
If a received message is available, this function returns the message in {\tt
m} and returns {\tt true}.

\item
{\tt passiveReceive(m)}
A call to this function blocks until either global termination
(of the current phase of the computation) is
detected, or a message sent to the calling process is available.
If termination is detected, the return result of this function is
{\tt false} and the value of {\tt m} is undefined.
If a received message is available, this function returns the message in {\tt
m} and returns {\tt true}.

\item {\tt hadFailureInLastPhase()}
This function returns a Boolean that indicates
  whether the last terminated phase involved a failed process.
\end{itemize}

\subsection{High Level Communication and Registry}

The High Level Communication (HLC) layer provides indirect,
anonymous communication among the players in a distributed
system.  It includes a number of local registries whose mutual
connectivity supports the full connectivity of the players
necessary for broadcast.  A player must sign-in at a local
registry, after which it can use the other operations provided by
the HLC layer to take part in the mechanism.
It provides the following functions:

%\vspace{-2mm}

% This function signs the calling player process in at the local
% registry {\tt r} and properly initializes the respective
% structures in both the registry {\tt r} and the calling player
% process.
% The player can start the first phase of the game \texttt{g} right
% after a successful return of a call to this function.

\begin{itemize}
\item {\tt signin(r, mech)} This function allows the calling player
  process to sign at the local registry {\tt r} so that it can take
  part in the mechanism \texttt{mech}.  The player can start the first
  phase of the mechanism \texttt{mech} right after a successful return
  of a call to this function.

\item {\tt signout(mech)}
This function terminates the participation of the calling player
process in the mechanism \texttt{mech}.

\item {\tt bsend(m)}
This function broadcasts the message {\tt m} to all registered
players in the game.

\item {\tt msend(m, T)}
This function multicasts the message {\tt m} to every player in
the target set {\tt T}.

\item {\tt receive(m, t)}, {\tt passiveReceive(m)}, {\tt hadFailureInLastPhase()}
These functions are the same as their homonyms in the BTTF layer.
\end{itemize}

Each local registry is responsible for processing the registrations of
the player processes according to the assumed registration criteria.
Also, it maintains for each implemented mechanism a corresponding
\textbf{locking policy}. Each such policy regulates the conditions
under which the player processes can receive messages sent to them or
can broadcast or multicast messages.  It is loaded each time a player
process successfully registers.  We shall return to this matter in the
next subsection and in Section \ref{subsec:nonGroves}.

\subsection{Player Process}
\label{subsec:player}

Specific applications are implemented using this top layer.
It is built on top of the HLC layer and is used to implement
specific actions of the players, in particular the computation
of the decisions and taxes. In our implementation of the
distributed mechanism design the following sequence of actions takes
place for each player $i$, where \texttt{flag} is a local Boolean variable.
By \texttt{termination loop} we mean here the statement
\begin{verbatim}
while (passiveReceive(m)) {
  process message m;
}
if (hadFailureInLastPhase()) {
 escalate detected failure;
}
\end{verbatim}
and by \texttt{inspect loop} we mean the statement
\begin{verbatim}
flag = false;
while (receive(m, 100)) {
  if (m is the pair (decision, tax scheme)) {
    flag = true;
    process message m;
  }
}
\end{verbatim}
where 100 is some arbitrary time-out in miliseconds.

The first loop allows the calling process to learn that all registered
player processes and the tax collector process have reached the same
phase in the distributed computation. At the end of the loop, if the
just terminated phase involved a process failure, this fact is
escalated in an application-dependent manner (e.g., by setting a flag
or calling an appropriate function) for proper handling.  In turn, the
inspection loop is used to determine whether another process has
already computed the decision and the tax scheme.

The details of the processing of each received message \texttt{m} depend on the context.

\begin{enumerate} \smallromani

 \item process $p_i$ representing player $i$ is created and assigned a globally unique name,
\label{label:1}

 \item $p_i$ obtains player $i$'s type,
\label{label:2}

 \item $p_i$ signs in at the local registry \texttt{r} in its region
   using the {\tt signin(r, mech)} call,
%    upon which all messages sent to
%    $p_i$ by processes representing other players are \emph{locked} and stored.
%    The lock prevents that $p_i$ can access these messages,
\label{label:3}

 \item if $p_i$ receives the confirmation of the registration (the call of {\tt signin(r, mech)} is successful),
it broadcasts player $i$'s type using the {\tt bsend()} function (and otherwise it terminates),
\label{label:4}

% \item the lock of $p_i$ is open so that $p_i$ can access all messages that were or will be sent to
%    it by  processes representing other players,
% \label{label:5}

\item $p_i$ performs the \texttt{termination loop}.  The corresponding
  \texttt{process m} statement in this loop consists of storing the
  type received from another registered player process.  When this
  loops properly ends (that is, when $p_i$ has received the types from all
  registered player processes and the global termination is detected)
  $p_i$ has a globally unique naming scheme at its disposal to refer
  to the registered player processes, and the number of registered
  players $n$ that equals the number of types it has received,
\label{label:6}

\item $p_i$ performs the \texttt{inspect loop}. If another process has
  already computed the decision and the tax scheme, \texttt{flag} will
  be set \texttt{true},
\label{label:7}

\item if \texttt{flag} is not \texttt{true}, $p_i$ computes the decision
  and the tax scheme of the players and multicasts using the {\tt
    msend()} function the decision and the tax scheme to the
  processes representing players who need to pay or receive taxes and the
  decision to the other processes. If $p_i$ needs to pay some tax 
  to the tax collector (respectively, to receive some tax return), 
it sends this information to the tax collector process using
  the {\tt msend()} function,
\label{label:8}

 \item $p_i$ performs the \texttt{termination loop},
\label{label:9}

\item when it properly ends and after $p_i$ receives from the tax
  collector process the information about the total amount of taxes
  (respectively, financial claims) the tax collector received, $p_i$
  performs the \texttt{termination loop} again and terminates.
\label{label:10}
\end{enumerate}

The last item refers to the tax collector process, described below,
with which all player processes jointly synchronize their computation
phases. One of the tasks of this process is to compute the aggregate
tax and communicate it to the player processes.  The details of the
tax scheme algorithm can be found in Appendix A.  The above
description of the player process assumes that the underlying
mechanism is \emph{simultaneous}. The corresponding locking policy,
loaded by the local registry, blocks the {\tt receive(m, t)} and {\tt
  passiveReceive(m)} functions of the $p_i$ process until it has
broadcast its type.

\subsection{Tax Collector Software Interface}

This layer is built on top of the HLC layer. It provides two functions
also available in the HLC layer, {\tt passiveReceive(m)} and
{\tt bsend(m)}, and two new functions, {\tt tsignin(r, mech)} and
{\tt  tsignout(mech)}, which are the counterparts of the {\tt signin(r, mech)} and
{\tt signout(mech)} functions of the HLC layer and which are used to deal
with the tax collector process registration.

\subsection{Tax Collector Process}
\label{subsec:tax}

This layer is built on top of the Tax Collector Software Interface
layer and is used to implement the actions of the tax
collector which is in charge of collecting players' taxes.
The following sequence of actions takes place for it:

\begin{enumerate} \smallromani
\item The tax collector process $ta$ representing the tax collector is
  created and assigned a globally unique name known to every player.
  It signs in at the local registry in his region using the {\tt
    tsignin(r, mech)} call (which always succeeds),
\label{label:t1}

\item $ta$ performs the \texttt{termination loop}, 
\label{label:t2}
  
\item $ta$ performs the \texttt{termination loop} again.  When it
  ends, the tax collector process has received all the taxes
  (respectively, financial claims) from the players. They are kept on
  a single account,
  
\item $ta$ broadcasts the information about the total amount on its
  single account to all players,

\item $ta$ performs the \texttt{termination loop} and terminates.
\label{label:t5}
\end{enumerate}

\subsection{Player GUI}

The interaction between the player (user) and the system is realized
in this interface.  The interaction is limited to the registration,
type submission and tax reception.

\subsection{The Initialization Phase}

Before a session of a given mechanism can start, a number of processes
and servers must be up and running. Every mechanism has its own
specific ``game server'' (Gserver) which is identified by a unique
\verb+<http-address>:<port-id>+. This URL will be publicized for
people who want to create and add new processes, such as player
processes, local registries, and tax collector processes, that will join the
current session of the mechanism. The Gserver ensures that no
player process enters the mechanism before the necessary local
registry and the tax collector processes are enrolled.
% What actually sits behind this address as a
% server is irrelevant for our purposes. (It can be for instance a
% single program running on a single machine, or it can be a distributed
% server running on several machines.) 

Each mechanism also has a second dedicated server, called the
``communication server'' (Cserver). Generally, the Cserver sits at a
different machine-and-port address than the Gserver and is used
internally only.  Its functionality is completely generic and
application independent, while the Gserver's functionality is
application dependent, which explains why these two servers are
separate.

The role of the Gserver is to perform mechanism specific
authentication necessary to make sure that only authorized processes
enter the game.  Once this is done, the Gserver gives the IP address
of its Cserver, together with an appropriate authorization key, to its
client process, which enables the latter to contact the Cserver and
authenticate itself for the proper communication services that it is
entitled to.

The role of the Cserver is to enable pairs of processes that run on
different machines to dynamically establish communication links
between themselves. When the Cserver receives a connection request, it
searches its tables to find a match for this request.  The details of
the rules by which two requests can match is beyond the scope of this
paper. In its simplest form, two requests to establish communication
by two parties match if each explicitly specifies the other party by
name or by category (i.e., player, registry, etc.). 

If no match is
possible for a new request, depending on the value of one of its
fields, it is either dropped, or retained in the internal tables of
the Cserver (for a possible match with a future request) and the
client is notified accordingly.
% The details of the rules by which two
% requests can match is beyond the scope of this paper. If no match is
% possible for a new request, depending on the value of its W field, it is
% either dropped, or retained in the internal tables of the Cserver and
% the client is notified accordingly. 
If a match is possible, then the Cserver replies to the two clients,
providing each with the address and the port-id of the other, which
subsequently proceed to establish their direct communication link
accordingly.

% The Cserver knows nothing about the
% application, mechanism rules, players, or registries.

To start the system for a new session of a mechanism, its Gserver is
first manually started up and its address is publicized. The first
thing that the Gserver does is to (create, if necessary and) connect
to a dedicated Cserver for this application and complete their
initialization by exchanging some identification information. The
Gserver is then ready to accept processes that wish to join this
session of the mechanism.

Once a process receives its successful authentification reply from the
Gserver, it completes its initialization by issuing some appropriate
requests to the Cserver to establish communication links with the
other processes in the application. The details of these requests
depend on the type of the process involved. We only mention them for
the player processes.

A player process contacts the Gserver and presents its credentials.
Once authenticated, the Gserver replies to its successful
authentication request supplying the player with the address of the
Cserver and the player's authorization key. It also sends in its reply
the \texttt{id} of the local registry where the player must register.
The player process then sends a request to the Cserver announcing its
readiness to communicate with its local registry. Once its
communication link with its local registry is established, the player
process signs in at the local registry (see item (\ref{label:3}) in
Subsection \ref{subsec:player}).  If the registration is confirmed
(item (\ref{label:4})), the local registry assigns the IP address of a
gateway to the player process. The player is then ready to take part
in this session of the mechanism.

Gateways play no functional role in the mechanism.  They exist only to
ensure the full connectivity of the network in its backbone of secure
hosts, thereby relieving the application layer of the concern for its
full connectivity. A gateway is a functionally empty process that acts
as a bridge and forwards the (control) messages of the BTTF layer.  We
assume sufficient redundancy (in the number of gateways and their
buddy sets\footnote{Every process must maintain a set of identifiers
  of $k\geq 0$ other processes in the system, called its {\em
    buddies}.  The buddies of processes are transparently used by the
  BTTF Wave algorithm to detect termination.  The only requirement on
  the buddy sets of processes is that they must collectively provide
  full reachability.})  to ensure full connectivity of the gateway
processes exists.
%
%
% A connection request <C, R, X, L, W, T> tells the Cserver that
% the client process C wishes to establish a communication link with a
% target process identified by X . The target process is either determined
% by C in its connection request as a specific process identifier, or it
% will be determined by the Cserver if X is a kind specification (e.g.,
% registry, player, tax collector, etc.), or one of the two special
% wild-card pattern symbols ' ?' or '*'.
%
%
% A connection request specifies the client C's local port R as where C
% expects to receive the Cserver's reply to this request. The connection
% request also specifies L as the local port where C expects its
% communication link with X to be established. The flag W specifies
% whether the Cserver should drop this request if it cannot be satisfied
% immediately (W = 0), retain the request until it is satisfied later with
% a single matching request (W = 1), or retain it until it is satisfied
% later, repeatedly, with multiple matching requests (W = 2). The lease
% time T specifies a length of time (> 0) for the Cserver to retain the
% request in its internal tables. The Cserver converts the lease time of a
% received request into an absolute expiration time by adding the current
% time to T before it further processes or keeps the request in its
% tables, and drops retained requests after their expiration time.
%
%

\subsection{Possible Realization}
\label{subsec:rea}

\begin{figure}[htbp]
%\centerline{\psfig{figure=Network1.eps,height=12cm}}
\includegraphics[height=12cm]{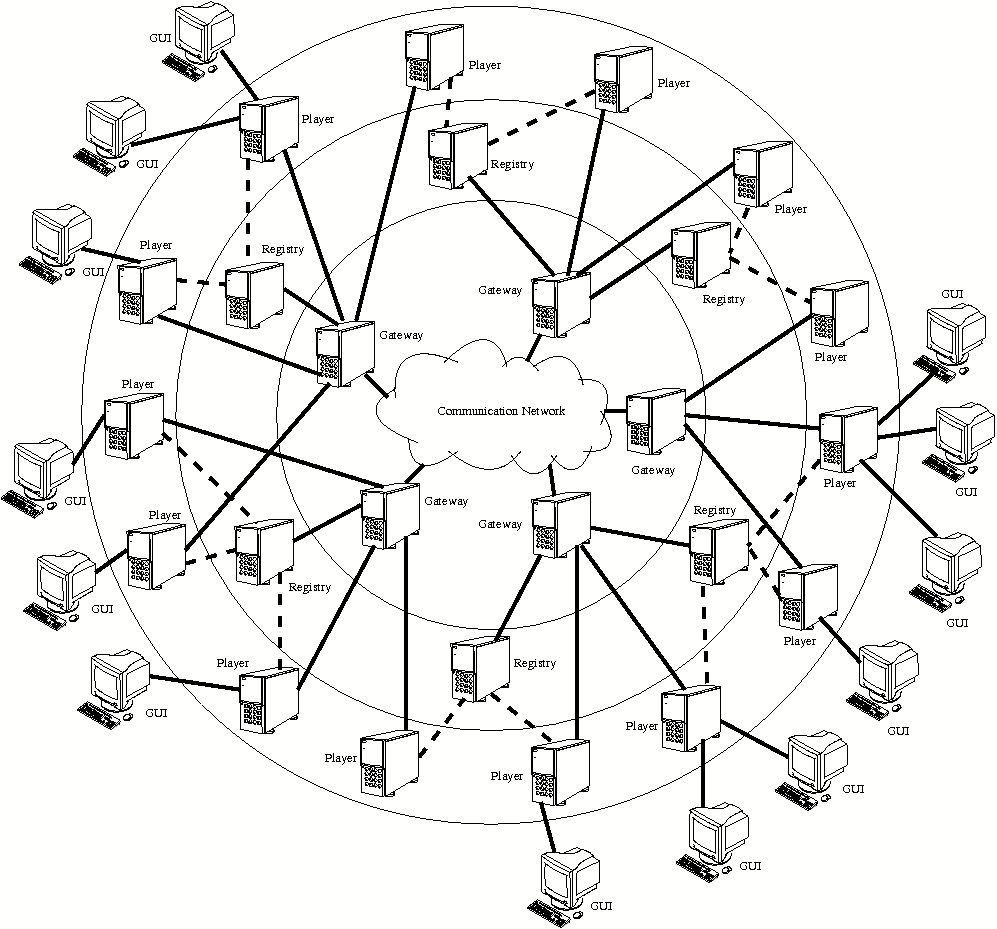}
\caption{\label{P} Possible realization of the platform}
\end{figure}

The architecture presented in this section allows multiple
alternative implementations, in each of which the constituents that
comprise each layer get allocated to run on a different set of hosts.
Figure \ref{P} shows an example mapping of constituents to `logical
hosts'. In any concrete implementation one or more such logical hosts
can represent the same actual physical host.

At the core of Figure \ref{P} lies a
\textbf{communications network}, represented by the cloud shape, that
interconnects a number of hosts to provide the functionality described
in the LLC layer in Subsection \ref{subsec:LLC}. The specific hosts
connected to this network that concern us are a set of \textbf{gateway
  hosts that run the BTTF and the HLC layers}.\footnote{By a gateway
  host we mean here a host that acts as an entry point to the
  communication network for player processes.}

The ring of hosts around the core in Figure \ref{P} contains the set of
\textbf{hosts that run the local registries}.  Every local registry
has a primary connection to a gateway host in the core.  Thus, the full
reachability of the gateway hosts in the core ensures full reachability
among local registries.

The next ring of hosts in Figure \ref{P} contains \textbf{hosts that run
player processes}.  Each player process establishes an initial link
(dotted lines) with a local registry (whose address it obtains from the
Gserver) to register.  As part of this registration process, its
local registry provides the address of a gateway host with which the
player process then establishes its primary communication link (solid
lines) for the rest of the game.

Finally, the outermost ring in Figure \ref{P} consists merely of
\textbf{computers that run GUI programs} that link to their respective
player processes.

\section{Extensions}
%\section{Security, Distributed Policing, Fault-tolerance and Repeated Mechanisms}
\label{sec:manip}

Let us discuss now some consequences of our design and some of its extensions.

\subsection{Security}

The  `ring structure' depicted in Figure \ref{P}
provides a multiple protection scheme against
manipulations by the players.  First, the assignment of a gateway host
to the player process, provided by the local registries, is done
dynamically. So there is no way for a player process to know
before-hand which host its local registry will propose as its gateway.

Next, the only messages that pass through a local registry are the
ones involving its locally registered players.  Likewise, the only
messages that pass through a player process are the ones originating
from or destined for that specific player.  This, as already mentioned
in Section \ref{sec:our}, automatically prevents manipulation by
player processes of messages originating from or destined for other
players.

Further, the end users have physical access only to the outermost
hosts that run the GUI programs, which severely restricts the range of
their potentially dangerous actions.  Finally, the separation of the
GUI programs from the player processes allows us to run the latter on
hosts to which end users do not have physical access.

Note also that the reliance on the registration process allows the
users to use the High Level Communication (HLC) layer in a `safe
mode'.  In such a mode the users can trust the security of the messages
they exchange through a `public' communications system, by relying on
the encryption of the messages using the public key cryptography.
This can be achieved, for example, by modifying the first call of
the \texttt{termination loop}, in action (\ref{label:6}) of Subsection
\ref{subsec:player}, so that it includes the collection of
public keys of the registered players.  Subsequent messages sent by
player processes can from that moment on be encrypted with recipients'
public keys.

In many applications it can be advisable to mask the identity of the
players, that is, to make it impossible to derive the identity of the
player from the name of the process representing it.  This can be
easily done by modifying action (\ref{label:4}) of Subsection
\ref{subsec:player}, so that once player process $p_i$ receives the
confirmation of the registration it also receives a new (globally
unique) name, securely generated by the local registry.

We do assume that the communications network, gateways and local
registries run on secure hosts.  The security issues involved here are
generic and independent of the properties and characteristics of any
specific mechanism in which the players may engage.  However, we do
not assume that player processes run on secure hosts, thus allow for the
possibility that they can be tampered with or tailor made, to
let their end users cheat.  In the next subsection we discuss how to
deal with this problem.

\subsection{Distributed Policing}

One possibility to tamper with the system consists of altering the
code of a player process so that it sends to some players a falsified
decision or a falsified tax scheme.  By \textbf{policing} we mean here
a sequence of actions that will lead to the exclusion of such
processes (that we call \emph{dishonest}).  The qualification
`distributed' refers to the fact that the policing is done by the
player processes themselves, without intervention of any central
authority.  Below we call a player process \emph{honest} if it
multicasts a true tax scheme.

The difficulty in implementing distributed policing lies in the fact
that dishonest processes may behave inconsistently. To resolve this
problem we use the registries which are assumed to be reliable.  We
then modify the sequence of actions of each player process so that it
always computes the decision and the tax scheme but sends them only to
its local registry.  The local registry then dispatches the decision
and the tax scheme on behalf of its sender to all player processes
mentioned in the tax scheme and the decision to the other processes.
As a trusted intermediary, the registry ensures that the same tax
scheme is sent to all player processes involved, and that no player
process can send more than one tax scheme in a single phase.

The resulting sequence of actions performed
by each player process $p_i$ is now as follows, where the new steps
are (\ref{label:a6})--(\ref{label:a8}):

 \begin{enumerate} \smallromani

  \item process $p_i$ representing player $i$ is created and assigned a globally unique name,
 \label{label:a1}

  \item $p_i$ obtains player $i$'s type,
 \label{label:a2}

  \item $p_i$ signs in at the local registry \texttt{r} in its region,
 \label{label:a3}

  \item if $p_i$ receives the confirmation of the registration
 it broadcasts player $i$'s type and otherwise it terminates,
 \label{label:a4}

 \item $p_i$ performs the \texttt{termination loop},
 \label{label:a5}

\item $p_i$ computes the decision and the tax scheme of the players
  and sends this information to its local registry, requesting the
  latter to dispatch this information, on its behalf, to all other player
  processes,
 \label{label:a6}

\item $p_i$ collects the decisions and the tax schemes dispatched by all other player
processes.  By comparing them with the true decision and tax scheme computed by
itself, $p_i$ identifies the set of honest player processes, $\emph{honest}_i$,
 \label{label:a7}

\item $p_i$ performs the \texttt{termination loop} and terminates.
 \label{label:a8}

\end{enumerate}

Note that upon termination each player process $p_j$ has the
same set $\emph{honest}_j$. This way all honest processes gain the
common knowledge of their own identities, which makes it possible for
them to `reconvene' in the case a falsified tax scheme was sent, or to
finalize the tax handling with the tax collector otherwise.

\subsection{Fault-tolerance}

Our system supports fault-tolerance on various levels. First, the
\texttt{llsend(m, r)} operation of the Lower Level Communication (LLC)
layer returns a Boolean value that indicates its success or failure,
so it provides a provision for recovery. 

Next, the BTTF algorithm from the BTTF layer detects persistent
process failures.  This means that should any process crash or
otherwise become non-responsive or unreachable at any time, the rest
of the processes in the distributed system can recover from this
failure and still reach termination.  BTTF uses probing to accomplish
this, nevertheless, the details of this protocol are not relevant in
the context of this paper.  The BTTF layer provides an inquiry
function to the application layer that enables it to determine whether
or not a process failed during the last phase whose termination was
detected.  Thus, the termination loop in our application can inform
the mechanism design level about the existence of failed processes,
which can then react appropriately to process failure.

Fault-tolerance on the mechanism design level has to address two
possibilities.

\begin{enumerate}
\item Some player processes crash before they broadcast their players’
types. To deal with this problem we implemented the following procedure.
In the first place the description of a given application is augmented
by a ‘deadline to react’ in the form of an absolute time, available to
each local registry. Recall that the actual broadcast of the type of
each player process $p_i$ representing player $i$ is actually carried out by
its local registry. When the deadline passes, each local registry makes
a list of the player processes that successfully registered at it and
from whom it did not receive their types. Each local registry then
informs other player processes about the exclusion of these ‘crashed’
player processes, so that the decision and the tax schemes can now be
computed only by the player processes that met the deadline.

\item Some player processes crash after they broadcast their players’
  types. This contingency is discovered by each player process by
  means of the \texttt{termi\-na\-tion loop} performed in step
  (\ref{label:a8}) of the last subsection.  How to deal with this
  problem is application dependent. 

In some applications the final
  decision and the tax scheme can be easily modified by all other
  player processes. For instance, in the problem of buying a path in a
  network discussed in Subsection \ref{subsec:Groves}, if a single
  player process crashes, the other player processes can always choose
  another path in the network. However, in other problems the whole
  decision process has to be restarted or aborted. This is for
  instance the case in the public project problem (also discussed in
  Subsection \ref{subsec:Groves}) when a ‘pivotal’ player process
  crashes, i.e., its tax changes the status of the project from
  carrying it out to dropping it. Then the sufficient level of funding
  is not available anymore.
\end{enumerate}

\subsection{Repeated mechanisms}

Finally, let us discuss the problem of implementing repeated
mechanisms that consist of several rounds of the same mechanism, each
round involving possibly different players. As a simple example, suppose
that we are dealing with the repeated rounds of a single item auction
and that the players are not allowed to purchase more than one item.

In each round different player processes may be admitted on the
basis of some information that needs to be maintained in the local
registries (like who were the winners in the previous rounds of the
single item auction). This means that the player processes should now
send the computed decision and the tax scheme also to the local
registries.  Then information maintained at the local registry allows
us to maintain a specific registration scheme used throughout the
repeated mechanism.

Apart from this modification the implementation is straightforward
thanks to the fact that all stages of the distributed computation
involving the player processes and the tax collector process have been
properly separated by the calls of the \texttt{termination loop}.
Consequently, the implementation of repeated mechanisms can be
obtained as a simple modification of the implementation discussed in
Section \ref{sec:details}.

Namely, we assume that the processes representing all
players are created first, that is assume that the step
(\ref{label:1}) of Subsection \ref{subsec:player} is performed once for all
players who will take part in some round of the considered mechanism.
Further we assume that the step (\ref{label:t1}) of Subsection
\ref{subsec:tax} is performed once by the tax collector process.
Then we simply iterate the rounds of the mechanism, where each round
consists of the steps (\ref{label:2})--(\ref{label:10}) of Subsection
\ref{subsec:player} for all player processes, and of the steps
(\ref{label:t2})--(\ref{label:t5}) of the tax collector process.

\section{Examples}
\label{sec:examples}

We used our distributed mechanism design system in a number of test
cases.  Each of them is implemented as an instantiation of the player
process layer described in Subsection \ref{subsec:player}.  

\subsection{Groves mechanisms}
\label{subsec:Groves}

We start by discussing briefly the implemented Groves mechanisms.
A more detailed description of them is given in Appendix C.

\paragraph*{Vickrey auction}
In Vickrey auction there is a single object for sale which is
allocated to the highest bidder who pays the second highest bid.
Our distributed implementation required addition of only 60 lines of code.

\paragraph*{Vickrey auction with redistribution}
We also implemented the proposal of \cite{Bai97} and \cite{Cav06} in which the
highest bidder redistributes some amounts from his payment to other
players. As shown in \cite{ACGM08} this Groves mechanism minimizes the
overall tax.

\paragraph*{Public projects}
The public project problem, see \cite{MWG95}, deals with the problem
of taking a joint decision concerning construction of a discrete public good,
for example a bridge. Each player needs to report his valuation from
the project when it takes place.  If the sum of the valuations
exceeds the cost of the project, the project takes place and each
player has to pay the same fraction of the cost. Otherwise the
project is canceled. We implemented for this problem the VCG mechanism.

\paragraph*{Unit demand auction}
In this auction there are multiple items offered for sale. We
assume that there are $n$ players and $m$ items and that each player
submits a valuation for each item.  The items should be allocated in
such a way that each player receives at most one of them
and the aggregated valuation is maximal.

To compute the decision in the VCG mechanism we used the Kuhn-Munkres
algorithm to compute the maximum weighted matching, where the weight
associated with the edge $(j,i)$ is the valuation for item $j$
reported by player $i$.  In our implementation we used the Java source
code available at \url{http://adn.cn/blog/article.asp?id=49}.  To
compute the tax for player $i$ in the VCG mechanism this algorithm is used
again, computing the maximum weighted matching with player $i$
excluded.

\paragraph*{Single-minded auction}
In this auction studied in \cite{LCS02}
there are $n$ players and $m$ items, with
each player only interested in a specific set of items
(which explains the name of the auction).  We implemented
the VCG mechanism for the situation in which each player $i$
is only interested in a consecutive sequence $a_i, \ldots, b_i$ of the
items $1, \ldots, m$, with $1 \leq a_i \leq b_i \leq m$.
The computations of the decision and of the
taxes involve computations of the maximum weighted matchings.

We note that in general, when each player is interested in an
arbitrary subset of the items, the computation of the decision in the
VCG mechanism is NP-hard, see \cite{LCS02}. 

\paragraph*{Buying a path in a network}
This Groves mechanism, due to \cite{NR01}, is concerned with
the problem of buying a path in a network.
Consider a communication network, modelled as a directed graph $G
  := (V,E)$ (with no self-cycles or parallel edges).  We assume that
  each edge $e \in E$ is owned by (a different) player $e$.
  We fix two vertices $s, t \in V$ and assume that for every edge $e$
  there is a $s-t$ path in $G$ with the edge $e$ removed.
  
  Each player (owning the edge) $e$ submits the cost $\theta_e$ of
  using $e$.  The central authority selects on the basis of players'
  submissions the shortest $s-t$ path in $G$.  This mechanism is not
  feasible and all players whose edges are selected submit financial
  claims to the tax collector.  The decision and the taxes are
  computed in a straightforward way, using Dijkstra's shortest path
  algorithm.

\subsection{Other mechanisms}
\label{subsec:nonGroves}

Additionally, we implemented the following two mechanisms that are not Groves mechanisms.

\paragraph*{Sequential Groves mechanisms}

In the original set up of the decision problem all players announce
their types independently.  In a modification studied in \cite{AE09}
the types are announced sequentially, in a random order.

Suppose that the random order is $1, \ldots, n$.  The crucial
difference between the customary set up and the one now considered is
that player \textit{i} knows the types announced by players $1,
\ldots, i-1$.  In \cite{AE09} it was shown that in Groves mechanisms
used for problems concerned with public projects players have then
other natural strategies than truth-telling (i.e., announcing their
true type). The nature of these strategies was clarified by showing
that when this mechanism is transformed to a simultaneous one, the
vector of the proposed strategies forms a Pareto optimal ex post Nash
equilibrium in a large class of so-called optimal strategies. In
particular the overall tax is minimized.

Sequential Groves mechanisms can be implemented by means of our
distributed mechanism system using the appropriate locking policy
loaded by the local registries.  This locking policy takes care that
when process $p_i$ receives the confirmation of the registration, it
includes its sequence number $j$ and information whether it represents
the last player (the latter is needed to use other optimal strategies
than truth-telling). Then the {\tt receive(m, t)} and {\tt
  passiveReceive(m)} functions of $p_i$ are partly blocked so that
only the messages sent by processes representing players with sequence
number $< j$ can be received, and the {\tt bsend()} function is
blocked until $p_i$ has received the types from these $j-1$ processes.
We used this approach to implement the specific sequential mechanism
for the public project problem introduced in \cite{AE09}.

% slightly modifying the player process layer,
% specifically items (\ref{label:4}) and (\ref{label:5}) in
% the sequence of actions described in Subsection \ref{subsec:player}
% to:

% \begin{itemize}

%  \item if process $p_i$ receives the confirmation of the registration,
% it includes its sequence number $j$ and information whether it represents the last player
% (the latter is needed to use other dominant
% strategies than truth-telling),

% \item the lock of $p_i$ is partly open so that $p_i$ can access all messages that were or will be sent to
%    it by processes representing players with sequence number $< j$,

% \item each process with sequence number $j$, where $j > 1$, counts the number of types it
%   received. When the count becomes $j-1$ it broadcasts the type,

% \item the lock of $p_i$ is then (completely) open.

% \end{itemize}

\paragraph*{Walker mechanism}

This mechanism, introduced in \cite{Wal81} deals with the continuous
public goods (for example grass area in a city).  Each player $i$ has
a utility function of the form $v_i(q) := b_i(q) - c_i(q)$. Here $q$
is the total amount of a continuous public good produced by the
players, $b_i(q)$ is the benefit for player $i$ from the amount of $q$
of public good, and $c_i(q)$ is the cost share player $i$ has to pay.

Each player $i$ reports a real number $x_i$, which is interpreted as the
amount of public good he agrees to produce.
Then he receives the payment (tax)
\[
t_i(x) := (x_{i+1} - x_{i-1}) \sum_{j = 1}^{n} x_j,
\]
where we interpret $n+1$ as 1 and $1-1$ as $n$, that is $i+1$ and
$i-1$ are the indices of the right-hand and left-hand neighbours of
player $i$ in a ring.
So $x = \sum_{j = 1}^{n} x_j$ is the total amount of public good produced
and the final utility for player $i$ is of the form
$u_i(x) := v_i(x) + t_i(x)$.

Walker mechanism is not a Groves mechanism and it implements the
decision not in dominant strategies but in an ex post Nash
equilibrium.\footnote{In an ex post Nash equilibrium no player can
  gain by deviating from his strategy even if he knows the types of
  the other players.}  Still, to implement it we again merely modified
the player process layer.  To test the implementation we used specific
functions $b_i$ and $c_i$.

\section{Conclusions and future work}
\label{sec:conc}

In this paper we discussed a design and implementation of a platform
that supports distributed mechanism design and that can be customized
to specific Internet-based applications.

We believe that the proposed platform clarifies how the design of
systems supporting distributed decision making through the Internet
can profit from sound and proven principles of software engineering,
such as separation of concerns and hierarchical design.  The discussed
platform is built as a sequence of layers. The lower layers provide
support for distributed computing, while the upper ones are concerned
only with the matters specific to mechanism design. Specific
Internet-based applications can be readily realized by creating a
backbone of local registries and selecting appropriate registration
details.

We found that the division of the software into layers resulted in a
flexible design that could be easily customized to specific mechanisms
proposed in the literature, such as (sequential) Groves mechanisms and
Walker mechanism, and to specific applications, such as various forms
of auctions.  For example, as already mentioned, our distributed
implementation of Vickrey auction required modification of a module of
only 60 lines of code.  Additionally, this layered architecture offers
a multi-level protection scheme against manipulations, distributed
policing and supports fault-tolerance.

We also provided evidence that software engineering in the area of
multiagent systems can profit from the techniques developed in the
area of distributed computing, for example broadcasting in an
environment with an unknown number of processes, distributed
termination and barrier synchronization.

In our work we have not dealt with the problem of false-name bids, see
\cite{Yok06}, that needs to be addressed anew in the context of
distributed implementations.  This is the subject of our current
research. Also, we plan to use our system to implement continuous double
auctions.

\section*{Acknowledgments}

The first two layers were implemented by Kees Blom who kindly modified
his implementation to our purposes.  Han Noot developed software for
message passing between internet-based parallel processes. The
applications to the unit demand and single-minded auctions were
suggested by Vangelis Markakis who also proposed the appropriate
algorithms. The work of Huiye Ma was funded by the NWO project
DIACoDeM, No 642.066.604.

\bibliography{/ufs/apt/bib/e,/ufs/apt/bib/sin02}
\bibliographystyle{abbrv}

\section*{Appendix A}

We explain here the details of the reduced tax scheme algorithm
mentioned in Section \ref{sec:our}.  Intuitively, this algorithm
determines given the tax vector $(t_1, \ldots, t_n)$ `who pays how much
to whom'.

We consider a list $L_{all}$ of players, each with his tax, and assume that -1 is
the identity of the tax collector.
First the players are divided into two lists, $L_{neg}$, consisting of players whose taxes are negative (i.e.,
those who should pay the taxes) and $L_{pos}$
consisting of players whose taxes are strictly positive (i.e., those
who should be paid). Players whose tax is 0 are omitted.

We start with the first player on the list $L_{neg}$, player
$A^0_{neg}$, and compare the absolute value of his tax, $|t^0_{neg}|$,
with the tax $t^0_{pos}$ of the first player on the list $L_{pos}$,
player $A^0_{pos}$.  If $|t^0_{neg}| < t^0_{pos}$, then player
$A^0_{neg}$ pays the amount $|t^0_{neg}|$ to player $A^0_{pos}$.  This
changes the tax of player $A^0_{pos}$ from $t^0_{pos}$ to
$t^0_{pos}+t^0_{neg}$. The process is now repeated with the next
player who should pay a tax, $A^1_{neg}$, and player $A^0_{pos}$.  If
$|t^0_{neg}| = t^0_{pos}$, the same happens but the process is
repeated with players $A^1_{neg}$ and $A^1_{pos}$.

If $|t^0_{neg}| > t^0_{pos}$, player $A^0_{neg}$ pays the amount
$t^0_{pos}$ to player $A^0_{pos}$. This changes the tax of player
$A^0_{neg}$ from $t^0_{neg}$ to $t^0_{neg}+t^0_{pos}$. The process is
now repeated with player $A^0_{neg}$ and the next unpaid player,
$A^1_{pos}$.

The loop stops when either all players with negative taxes paid or all
players with strictly positive taxes were paid.  If the mechanism is
not budget balanced, upon termination either each player $l$ who at
this moment still needs to pay taxes (his tax is negative) pays it to
the tax collector ---this is represented by the triple
$(l,|t^l_{neg}|,-1)$--- or each player $l$ who at this moment still
should be paid (his tax is strictly positive) sends his financial
claim to the tax collector ---this is represented by the triple
$(l,-t^l_{pos},-1)$.

The pseudo-code of the algorithm is given in Figure \ref{fig:tax}.

\vspace*{-5mm}

\begin{figure}[htbp]
\small

\vspace*{-10mm}

\hrule height0.8pt\vspace{5.8pt}
{
%\linesnumbered
\begin{algorithm}[H]
\SetLine
$L_{all}$ is the list of players\;
$all$ is the length of the list $L_{all}$\;
$A^i$ is the $(i+1)$st player in the list $L_{all}$\;
$t^i$ is the tax of player $A^i$\;
$tax$ is the list representing the computed tax scheme\;
$tax = nil$; $neg = 0$; $pos = 0$\;
% $tax = nil$\;
% $neg = 0$\;
% $pos = 0$\;
\For{$i=0$ \textbf{to} $all$}{
\eIf{$t^i < 0$}{
append $A^i$ to the list $L_{neg}$\;
$neg = neg +1$\;
}{
\If{$t^i > 0$}{
append $A^i$ to the list $L_{pos}$\;
$pos = pos +1$\;
}}
% \If{$t^i < 0$}{
% append $A^i$ to the list $L_{neg}$\;
% $neg = neg +1$\;
% }
% \If{$t^i > 0$}{
% append $A^i$ to the list $L_{pos}$\;
% $pos = pos +1$\;
% }

} %endfor
%----------
$A^j_{neg}$ is the $(j+1)$st player in the list $L_{neg}$\;
$t^j_{neg}$ is the tax of player $A^j_{neg}$\;
$A^k_{pos}$ is the $(k+1)$st player in the list $L_{pos}$\;
$t^k_{pos}$ is the tax of player $A^k_{pos}$\;
$j = 0$\;
$k = 0$\;
\While{$j < neg$ \textbf{and} $k < pos$}{
        \eIf{$|t^j_{neg}| \leq t^k_{pos}$}{
        append $(j, |t^j_{neg}|, k)$ to the list $tax$; \%  $A^j_{neg}$ pays the amount $|t^j_{neg}|$ to  $A^k_{pos}$\
        $t^k_{pos} = t^k_{pos} + t^j_{neg}$\;
        $j = j+1$\;
        \If{$|t^k_{pos}|==0$}{$k=k+1$}}{
        append $(j, t^j_{pos}, k)$ to the list $tax$; \%  $A^j_{neg}$ pays the amount $t^k_{pos}$ to  $A^k_{pos}$\
        $t^j_{neg} = t^j_{neg} + t^k_{pos}$\;
        $k = k+1$\;}
}%endwhile
\If{$j < neg$}{
\For{$l=j$ \textbf{to} $neg-1$}{
append $(l, |t^l_{neg}|, -1)$ to the list $tax$; \%  $A^l_{neg}$ pays the amount $|t^l_{neg}|$ to the tax collector}}
\If{$k < pos$}{
\For{$l=k$ \textbf{to} $pos-1$}{
append $(l, - t^l_{pos}, -1)$ to the list $tax$; \%  $A^l_{neg}$ claims the amount $t^l_{pos}$ from the tax collector}}
\end{algorithm}

}
\vspace{5pt}\hrule height 0.8pt
%\setlength{\abovecaptionskip}{15.8pt}
%\vspace{-3mm}

\caption{The algorithm to compute reduced tax scheme \label{fig:tax}}
\end{figure}

\normalsize

\newpage

\medskip

%%\begin{quote}

\section*{Appendix B}

\begin{figure}[htbp]
 \begin{center} %\ \setlength{\epsfxsize}{3cm}
% \epsfbox{pc1-registry5002-2.epsf}
\includegraphics[height=5cm]{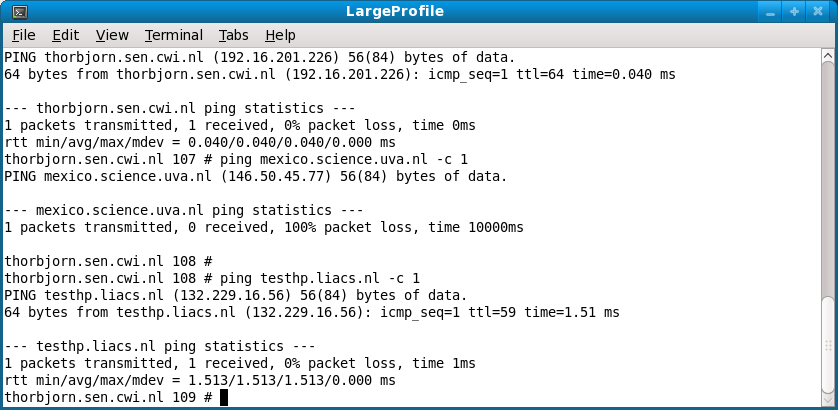}
\end{center}
 \caption{Three PCs at three locations.\label{fig:ping}}
 \end{figure}

In this appendix, we present an example interaction with the platform.
This test was run on three PCs, thorbjorn.sen.cwi.nl, testhp.liacs.nl
and mexico.science.uva.nl, located respectively at CWI, University of
Amsterdam and University of Leiden. Their IP addresses are shown in
Figure \ref{fig:ping}.
We assume that each player chooses from the pull down menu a
single-minded auction, discussed in Section \ref{sec:examples}. We
consider a specific instance with
\begin{itemize}
 \item 3 items for sale,

 \item 2 local registries, thorbjorn.sen.cwi.nl:8802 and  mexico.science.uva.nl:8803; we use time as a deadline for registration,

 \item 1 tax collector thorbjorn.sen.cwi.nl:8801,
   
 \item 6 players: player testhp.liacs.nl:8806--8809 and
   mexico.science.uva.nl:8804--8805 who register randomly in two local
   registries. Among these six players, we arrange that one player,
   testhp.liacs.nl:8809, is a latecomer who registers after the
   deadline, and as a result is not accepted to participate in the
   auction,
  
\item the following players' bids: 
\II

testhp.liacs.nl:8806: 20(1,2),
  testhp.liacs.nl:8807: 50(3), 

testhp.liacs.nl:8808: 32:(2),
   mexico.science.uva.nl:8805: 60(2,3), 

 mexico.science.uva.nl:8804: 19(1),
\II

  that is player testhp.liacs.nl:8806 bids 20 for the bundle (1,2),
  etc.,

 \item 3 PCs: PC1 (thorbjorn.sen.cwi.nl), PC2 (testhp.liacs.nl), and PC3 (mexico.science.uva.nl); PC1 runs the tax collector (thorbjorn.sen.cwi.nl:8801) and the
  local registry (thorbjorn.sen.cwi.nl:8802); PC2 runs four players (testhp.liacs.nl:8806--8809), and PC3
  runs the local registry (mexico.science.uva.nl:8803) and two players (mexico.science.uva.nl:8804--8805).
\end{itemize}

In this example the generated allocation is 
\[
\mbox{(3:testhp.liacs.nl:8807,28), (2:testhp.liacs.nl:8808,10), (1: mexico.science.uva.nl:8805,0), }
\]
that is item 3 is sold to player testhp.liacs.nl:8807 who pays for it to the tax
collector 28, etc.

The interaction with the system is presented in Figures \ref{pc1-atdeadline100}--\ref{pc3-endA} below.
It consists of four phases.

\medskip

\noindent
\textbf{Phase 1}.

\begin{figure}[htbp]
 \begin{center} %\ \setlength{\epsfxsize}{3cm}
% \epsfbox{pc1-bank5001-1.epsf}
\includegraphics[height=5cm]{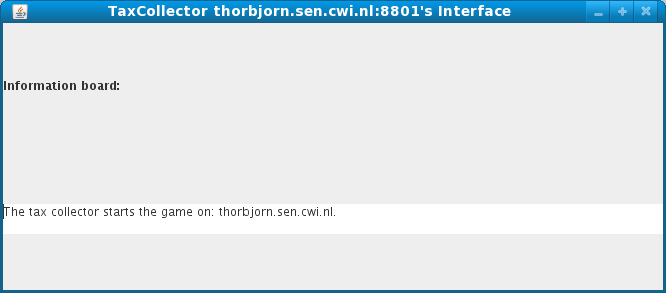}
 \end{center}
 \caption{Phase 1: tax collector thorbjorn.sen.cwi.nl:8801 on PC1 (thorbjorn.sen.cwi.nl). \label{pc1-atdeadline1000}}
 \end{figure}

This phase consists of the registration process.  Selected elements of
it are presented in Figures
\ref{pc1-atdeadline1000}--\ref{pc3-endA}.  Figures
\ref{pc1-atdeadline1000}  and \ref{pc1-atdeadline100} depict start of
the tax collector thorbjorn.sen.cwi.nl:8801 and the local registry
thorbjorn.sen.cwi.nl:8802 on PC1.  The information board of the latter
shows that players testhp.liacs.nl:8806 and mexico.science.uva.nl:8805
registered at this local registry.

The subsequent two Figures, \ref{pc2-atdeadlineC} and
\ref{pc3-atdeadlineA}, show successful registration of players
testhp.liacs.nl:8806 and mexico.science.uva.nl:8804.
% on thorbjorn.sen.cwi.nl:8802 
% and player testhp.liacs.nl:8809 trying to
% register.
%  on mexico.science.uva.nl:8803 .
% , while Figures
% \ref{pc3-atdeadlineA}--\ref{pc3-atdeadlineB} show (processes
% representing) players mexico.science.uva.nl:8804 and
% mexico.science.uva.nl:8805 trying to register on
% thorbjorn.sen.cwi.nl:8802.
Player testhp.liacs.nl:8809 was late in his attempt to register in the
local registry mexico.science.uva.nl:8803 and his registration was
rejected, as shown on Figures \ref{pc2-latecomer} and
\ref{pc3-lateregistration}. The last figure also shows which players
successfully registered at mexico.science.uva.nl:8803.

\medskip

\noindent
\textbf{Phases 2-4}.

The second phase consists of the type submission by players whose
registration has been accepted.  The third phase consists of the
computation of the tax scheme by the registered players, its
multicasting to other players and (in case of budget unbalanced
mechanism) sending payments (respectively, financial claims) of the
remaining taxes (respectively, claims) to the tax collector.  The
fourth phase consists of receiving by the registered players
information from the tax collector about the aggregate tax
(respectively, claim) received by it.

These three phases of the running example are depicted in Figures
\ref{pc2-endE} and \ref{pc3-endA} showing the information boards of
players testhp.liacs.nl:8808 and mexico.science.uva.nl:8804. In this
example, in phase 3, the tax scheme was computed by every player who
subsequently broadcast the computation result,
((testhp.liacs.nl:8807,-1,28),(testhp.liacs.nl:8808,-1,10)), to all
the players registered in the auction.

 \begin{figure}[htbp]
 \begin{center} %\ \setlength{\epsfxsize}{3cm}
% \epsfbox{pc1-registry5002-2.epsf}
\includegraphics[height=5cm]{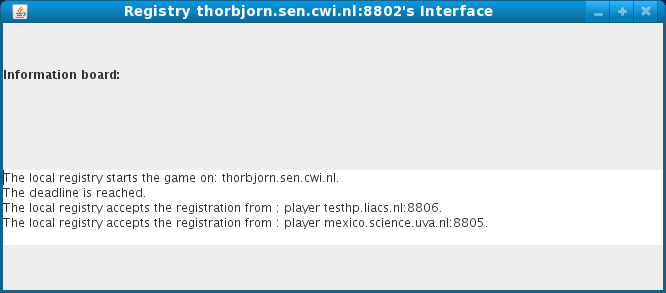}
\end{center}
 \caption{Phase 1: local registry thorbjorn.sen.cwi.nl:8802 on PC1 (thorbjorn.sen.cwi.nl). \label{pc1-atdeadline100}}
 \end{figure}

 \begin{figure}[htbp]
 \begin{center} %\ \setlength{\epsfxsize}{3cm}
% \epsfbox{pc2-player5006-2.epsf}
\includegraphics[height=5cm]{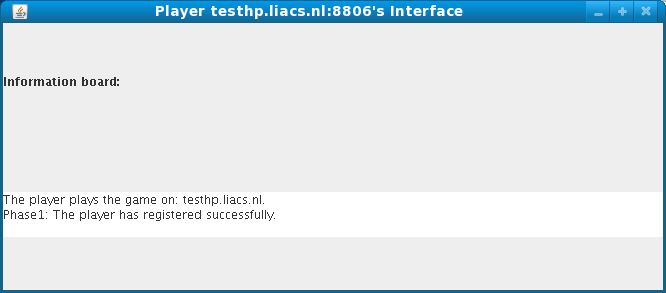}
 \end{center}
 \caption{Phase 1: player testhp.liacs.nl:8806 on PC2 (testhp.liacs.nl). \label{pc2-atdeadlineC}}
 \end{figure}

%  \begin{figure}[htbp]
%  \begin{center} %\ \setlength{\epsfxsize}{3cm}
% % \epsfbox{pc2-player5007-2.epsf}
% \includegraphics[height=5cm]{F7.png}
%  \end{center}
%  \caption{Phase 1: player testhp.liacs.nl:8807 on PC2 (testhp.liacs.nl). \label{pc2-atdeadlineD}}
%  \end{figure}

%  \begin{figure}[htbp]
%  \begin{center} %\ \setlength{\epsfxsize}{3cm}
% % \epsfbox{pc2-player5008-2.epsf}
% \includegraphics[height=5cm]{F8.png}
%  \end{center}
%  \caption{Phase 1: player testhp.liacs.nl:8808 on PC2 (testhp.liacs.nl). \label{pc2-atdeadlineE}}
%  \end{figure}

 \begin{figure}[htbp]
 \begin{center} %\ \setlength{\epsfxsize}{3cm}
% \epsfbox{pc3-player5004-2.epsf}
\includegraphics[height=5cm]{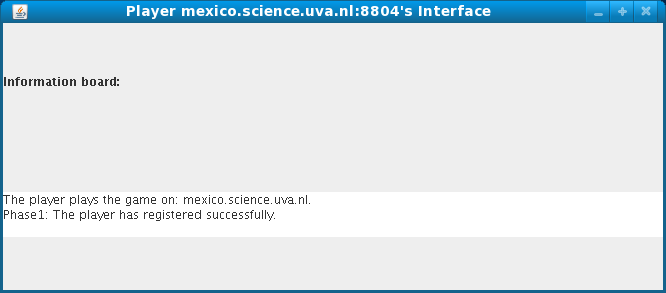}
 \end{center}
 \caption{Phase 1: player  mexico.science.uva.nl:8804 on PC3 (mexico.science.uva.nl). \label{pc3-atdeadlineA}}
 \end{figure}

%  \begin{figure}[htbp]
%  \begin{center} %\ \setlength{\epsfxsize}{3cm}
% % \epsfbox{pc3-player5005-2.epsf}
% \includegraphics[height=5cm]{F10.png}
%  \end{center}
%  \caption{Phase 1: player  mexico.science.uva.nl:8805 on PC3 (mexico.science.uva.nl). \label{pc3-atdeadlineB}}
%  \end{figure}

 \begin{figure}[htbp]
 \begin{center} %\ \setlength{\epsfxsize}{3cm}
% \epsfbox{pc2-player5009-late.epsf}
\includegraphics[height=5cm]{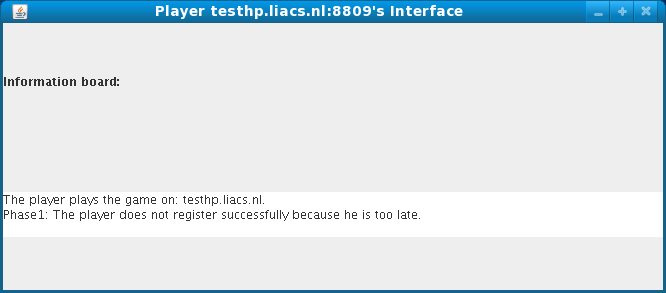}
 \end{center}
 \caption{Phase 1: latecomer player testhp.liacs.nl:8809 on PC2 (testhp.liacs.nl). \label{pc2-latecomer}}
 \end{figure}

 \begin{figure}[htbp]
 \begin{center} %\ \setlength{\epsfxsize}{3cm}
% \epsfbox{pc3-registry5003-2.epsf}
\includegraphics[height=5cm]{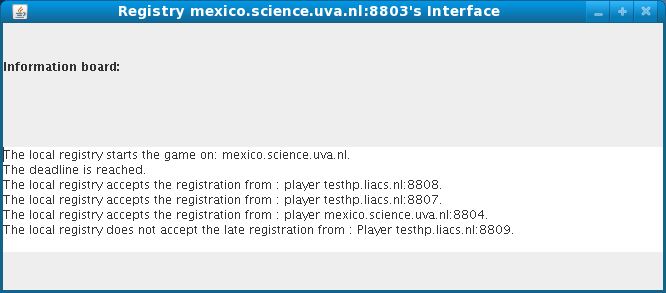}
 \end{center}
 \caption{Phase 1: local registry  mexico.science.uva.nl:8803 with late registration on PC3 (mexico.science.uva.nl). \label{pc3-lateregistration}}
 \end{figure}

%  \begin{figure}[htbp]
%  \begin{center} %\ \setlength{\epsfxsize}{3cm}
% % \epsfbox{pc2-player5006-4.epsf}
% \includegraphics[height=5cm]{F13.png}
%  \end{center}

%  \caption{Phases 2-4: player testhp.liacs.nl:8806 on PC2 (testhp.liacs.nl). \label{pc2-endC}}
%  \end{figure}

%  \begin{figure}[htbp]
%  \begin{center} %\ \setlength{\epsfxsize}{4cm}
% % \epsfbox{pc2-player5007-4.epsf}
% \includegraphics[height=5cm]{F14.png}
%  \end{center}

%  \caption{Phases 2-4: player testhp.liacs.nl:8807 on PC2 (testhp.liacs.nl). \label{pc2-endD}}
%  \end{figure}

\newpage

 \begin{figure}[htbp]
 \begin{center} %\ \setlength{\epsfxsize}{4cm}
% \epsfbox{pc2-player5008-4.epsf}
\includegraphics[height=5cm]{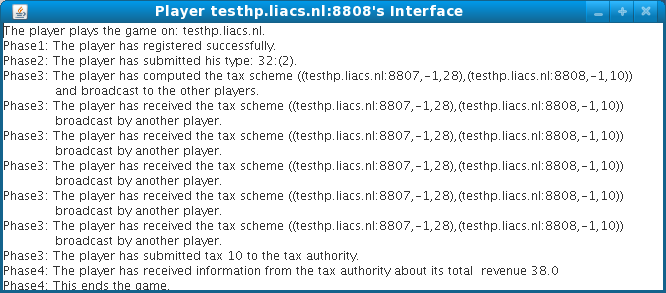}
 \end{center}

 \caption{Phases 2-4: player testhp.liacs.nl:8808 on PC2 (testhp.liacs.nl). \label{pc2-endE}}
 \end{figure}

 \begin{figure}[htbp]
 \begin{center} %\ \setlength{\epsfxsize}{4cm}
% \epsfbox{pc3-player5004-4.epsf}
\includegraphics[height=5cm]{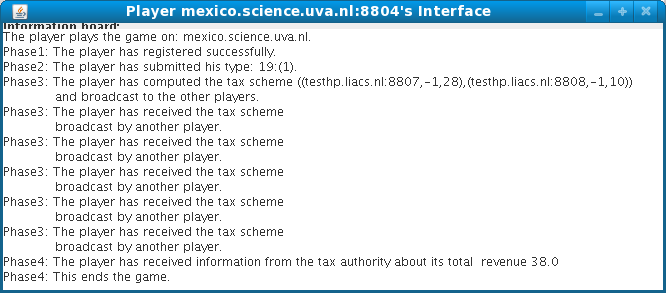}
 \end{center}
 \caption{Phases 2-4: player  mexico.science.uva.nl:8804 on PC3 (mexico.science.uva.nl). \label{pc3-endA}}
 \end{figure}

%  \begin{figure}[htbp]
%  \begin{center} %\ \setlength{\epsfxsize}{4cm}
% % \epsfbox{pc3-player5005-4.epsf}
% \includegraphics[height=5cm]{F17.png}
%  \end{center}
%  \caption{Phases 2-4: player  mexico.science.uva.nl:8805 on PC3 (mexico.science.uva.nl). \label{pc3-endB}}
%  \end{figure}

\section*{Appendix C}

We discuss here in more detail the examples of Groves mechanisms
implemented using our platform by explaining how each of them can be
represented as a decision problem. In each case it is straightforward
to check that the considered decision function is efficient.
Consequently in each case Groves theorem applies.

\paragraph*{Vickrey auction}
We model Vickrey auction as the following decision problem
$(D,\Theta_1,\ldots,$ $\Theta_n, v_1,\ldots,v_n,f)$:

\begin{itemize}

\item $D = \{1, \ldots, n\}$,

 \item each $\Theta_i$ is the set $\mathbb{R}_+$ of non-negative reals;
$\theta_i \in \Theta_i$ is player $i$'s valuation of the object,

\item
$
        v_i(d, \theta_i) :=
        \left\{
        \begin{array}{l@{\extracolsep{3mm}}l}
        \theta_i   & \mathrm{if}\  d = i \\
        0      & \mathrm{otherwise}
        \end{array}
        \right.
$

\item
$
f(\theta) := i,
$

where $\theta_i = \max_{j \in [1..n]} \theta_j$ and\footnote{In case of a tie we allocate the
object to the player with the highest index.} $\forall j \in [i+1..n] \: \theta_j < \theta_i$.
\end{itemize}

Here decision $d \in D$ indicates to which player the object is sold.
Below, given a sequence $s$ of reals we denote by $[s]_k$ the $k$th largest
element in this sequence. For example, for $\theta = (1,5,2,3,2)$ we have
$[\theta_{-2}]_2 = 2$ since $\theta_{-2} = (1,2,3,2)$.

The payments (taxes) in Vickrey auction are realized by applying the VCG mechanism, which yields
\[
        t^V_i(\theta) :=
        \left\{
        \begin{array}{l@{\extracolsep{3mm}}l}
        - [\theta]_{2}   & \mathrm{if}\  f(\theta) = i \\
        0      & \mathrm{otherwise}
        \end{array}
        \right.
\]

\paragraph*{Vickrey auction with redistribution}
To formalize the redistribution scheme of \cite{Cav06} in our framework
we combine each tax $t^V_i$ in Vickrey auction 
with the following function $h_i$ (to ensure that it is well-defined we need
to assume that $n \geq 3$):
\[
h_i(\theta_{-i}) := \frac{[\theta_{-i}]_2}{n}
\]
that is, by using
\[
t_i(\theta) := t^V_i(\theta) + h_i(\theta_{-i}).
\]
One can easily show (see \cite{Cav06}) that this yields a feasible Groves mechanism.

Let, given the sequence $\theta$ of submitted bids (types), $\pi$ be the permutation of
$1, \ldots, n$ such that $\theta_{\pi(i)} = [\theta]_i$ for $i \in [1..n]$ (where we break the ties by selecting
players with the higher index first). So the $i$th highest bid is by player $\pi(i)$ and the object is sold
to player $\pi(1)$.
Then

\begin{itemize}
\item $[\theta_{-i}]_{2} = [\theta]_{3}$ for $i \in \{\pi(1), \pi(2)\}$,

\item $[\theta_{-i}]_{2} = [\theta]_{2}$ for $i \in \{\pi(3), \ldots, \pi(n)\}$,
\end{itemize}
so the above Groves mechanism boils down to the following payments by player $\pi(1)$:

\begin{itemize}
\item $\frac{[\theta]_{3}}{n}$ to player $\pi(2)$,
\item $\frac{[\theta]_{2}}{n}$ to players $\pi(3), \ldots, \pi(n)$,
\item $[\theta]_2 - \frac{2}{n}[\theta]_{3} - \frac{n-2}{n} [\theta]_{2} = \frac{2}{n}([\theta]_2 - [\theta]_{3})$ to the tax collector,
\end{itemize}
which is how it was originally defined in \cite{Cav06}.

\paragraph*{Public projects}
\label{subsec:public}
The public problem corresponds to the following decision problem:

\begin{itemize}
 \item $D=\{0,1\}$ (reflecting whether a project is cancelled or takes place),

 \item $\Theta_i = \mathbb{R}_+$,

 \item $v_i(d,\theta_i):=d(\theta_i-\frac{c}{n})$,
 \item
$
f(\theta) := \left\{ \begin{array}{ll}
1 & \textrm{if $\sum_{i=1}^n \theta_i\ge c$}\\
0 & \textrm{otherwise}
\end{array} \right.
$
\end{itemize}

Here \textit{c} is the cost of the project. If the project
takes place, $\frac{c}{n}$ is the cost share of the project for each
player. We ensure that the decision rule is publicly
known by initially providing each player with the cost of the project
$c$.

\paragraph*{Unit demand auction}
We assume that there are $n$ players and $m$ items and that each player
submits a valuation for each item.  The items should be allocated in
such a way that each player receives at most one of them
and the aggregated valuation is maximal.
The unit demand auction with $n$ players and $m$ items 
can be modelled as the following decision problem:
\begin{itemize}

\item
$D  =  \{f :  \{1, \ldots, n\} \myra \{0,1, \ldots, m\} \mid \fa j \in \{1, \ldots, m\} \te^{\leq 1} i \in \{1, \ldots, n\} \: f(i) = j\}$,

where $\te^{\leq 1}$ stands for `there exists at most one', i.e., each decision is a 1-1 allocation of items to players,

\item $\Theta_i = \mathbb{R}^{m}_{+}$; $(\theta_{i, 1}, \ldots, \theta_{i, m}) \in \Theta_i$ is a vector of
player $i$'s valuations of the items for sale,

 \item
$
v_i(d,\theta_i) := \left\{ \begin{array}{ll}
\theta_{i, j} & \textrm{if $d(j) = i$} \\
0 & \textrm{if $\neg \exists j \: d(j) = i$}
\end{array} \right.
$

\item
$
f(\theta') := \textrm{$d$ for which $\sum_{j \in dom(d)} \theta'_{d(j), j}$ is maximal}.
$
\end{itemize}

\paragraph*{Single-minded auction}
We assume that there are $n$ players and $m$ items and each player $i$
is only interested in a consecutive sequence $a_i, \ldots, b_i$ of the
items $1, \ldots, m$, with $1 \leq a_i \leq b_i \leq m$.

We model this as the following decision problem:

\begin{itemize}

\item  $D = \{f \mid f: A \rightarrow \{1, \ldots, n\}, \ A \subseteq \{1, \ldots, m\}\}$,

\item $\Theta_i = \mathbb{R}_{+}$; $\theta_i \in \Theta_i$ is player
  $i$'s valuation for the sequence $a_i, \ldots, b_i$ of the items,

\item
$
v_i(d,\theta_i) := \left\{ \begin{array}{ll}
\theta_{i} & \textrm{if $d(j) = i$ for all $j \in [a_i, \ldots, b_i]$} \\
0 & \textrm{otherwise}
\end{array} \right.
$

\item
$
f(\theta') := \textrm{$d$ for which $\sum_{i: d([a_i, \ldots, b_i]) = \{i\}} \theta'_{i}$ is maximal},
$

where $d([a_i, \ldots, b_i]) = \{d(j) \mid j \in [a_i, \ldots, b_i]\}$.

\end{itemize}

So, given an allocation $f \in D$ the goods in the set $\{k \mid f(k) = j\}$
are allocated to player $j$.
Note that alternatively $f$ can be defined by:
\[
f(\theta') := \textrm{$d$ for which $\sum_{i = 1}^{n} v_i(d, \theta'_{i})$ is maximal}.
\]

\paragraph*{Buying a path in a network}
We also implemented the Groves mechanism concerned with
the problem of buying a path in a network, introduced in \cite{NR01}.
We consider a communication network, modelled as a directed graph $G
  := (V,E)$ (with no self-cycles or parallel edges).  We assume that
  each edge $e \in E$ is owned by (a different) player $e$.
  We fix two vertices $s, t \in V$ and assume that for every edge $e$
  there is a $s-t$ path in $G$ with the edge $e$ removed.

Each player (owning the edge) $e$ submits the cost $\theta_e$ of using $e$.
The central authority selects on the basis of players' submissions the shortest $s-t$ path in $G$.

This problem can be modelled as the following decision problem, where we denote the edges by
letters $i,j$:

\begin{itemize}
 \item $D = \{p \mid p \mbox{ is a $s-t$ path in $G$}\}$,

 \item each $\Theta_i$ is  $\mathbb{R}_+$; $\theta_i$ is player's $i$ incurred cost if the edge $i$ is used in the selected path,

 \item $
     v_i(p, \theta_i) :=
        \left\{
        \begin{array}{l@{\extracolsep{3mm}}l}
        - \theta_i   & \mathrm{if}\  i \in p \\
        0      & \mathrm{otherwise}
        \end{array}
        \right.
$
 \item
$f(\theta) := p$, where $p$ is the shortest $s-t$ path in $G$, where the cost of each edge $i$ is $\theta_i$.

In the case of multiple shortest paths we select, say, the one that is alphabetically first.
\end{itemize}

One can show that the following taxes (that are payments to the
players) constitute a Groves mechanism (see \cite{NR01}):
\[
t_i(\theta) =
        \left\{
        \begin{array}{l@{\extracolsep{3mm}}l}
        cost(p_1) - cost(p)  + \theta_i & \mathrm{if}\  i \in p \\
        0      & \mathrm{otherwise}
        \end{array}
        \right.
\]
where we abbreviate $\sum_{j \in p} \theta_j$ to $cost(p)$,
$p$ is the shortest $s-t$ path in $G$ selected by the
decision rule $f$ (that is, $f(\theta) = p$)
and $p_1$ is a shortest $s-t$ path in $G$ that does
not include the edge $i$.

\end{document}